\documentclass[UTF8,a4paper]{article}
\usepackage[fleqn]{amsmath}
\usepackage{amsfonts,amssymb,amsthm}
\usepackage{siunitx}
\usepackage{geometry}
\usepackage{xcolor}
\usepackage{hyperref}
\usepackage{mathtools}
\usepackage{mathrsfs}
\usepackage{upgreek}

\hypersetup{colorlinks,bookmarks=true,bookmarksnumbered=true,%
    pdfstartview=FitH,linkcolor=black,anchorcolor=blue,%
    citecolor=green}
\title{Are conservation laws independent of field equations?}
\author{Liu Changli \thanks{LiucL78@qq.com . }}

\date{}

\begin{document}
    \maketitle

    \begin{abstract}
            The charge conservation law ($\partial_{\alpha} J^\alpha =0$) usually is considered 
            as a corollary of Maxwell's equations and is not independent of Maxwell's field equations. 
            A circular reasoning, however,  is found in the derivation.
            A similar fallacy exists in the matter source's conservation law ($\nabla^{\mu}T_{\mu\nu}=0$)
            and Einstein's field equations. Therefore, the source's conservation laws are independent of Einstein's field equations.
    \end{abstract}

    \maketitle


\section{Introduction}

Let us illustrate viewpoints of the current theoretical physics literatures about the relationship between 
 conservation laws of sources and field equations.

For electromagnetic fields, the charge conservation law is regarded as a corollary of Maxwell's equations and is not independent. 
The derivation is very simple. Assuming Maxwell's equations are correct, and then
\begin{equation}
    \partial^\nu F_{\mu\nu} = J_\mu \quad \Rightarrow \quad
    0= \partial^\mu\partial^\nu F_{\mu\nu} = \partial^\mu J_\mu
    \quad \Rightarrow \quad \partial^\mu J_\mu=0 .
\end{equation}
The antisymmetric property of the electromagnetic field tensor $F_{\mu\nu}$ is used in the last step above.
The literatures supporting this view are \cite{feynman2, griffiths, jackson, LL1980, pauli2000} and so on.

For gravitational field, the sources conservation laws ($\nabla^a T_{ab}=0 $)are thought as a consequence of 
Einstein field equations($G_{ab}=8\pi T_{ab}$) , and  are not independent. The derivation are as following,
\begin{equation}
    0=\nabla^a G_{ab} = 8\pi \nabla^a T_{ab}\quad \Rightarrow \quad \nabla^a T_{ab}=0 .
\end{equation}
The literatures supporting this view are\cite{mtw1973, pais, poisson} and so on.

It, however, is \emph{not true}.  


\section{Existence of solutions for differential equations}

We roughly establish the axiomatic system of differential equations.

\begin{table}[htb]
    \centering
    \caption{} \label{chfd:tab-diff-axiom}
    \begin{tabular}{|*2{c|}}
        \hline
        Levels & contents \\ \hline
        Level 0 & the differential equations themselves   \\ \hline
        Level 1 & the existence theorem of solutions of differential equations \\ \hline
        Level 2 & all other theorems excluding "the existence theorem"  \\ \hline
    \end{tabular}
\end{table}

"Level 0" is an axiom level. In this case, only the differential equations themselves, such as the first order ordinary differential equation, Maxwell's equations, Einstein's field equations, and so on.

"Level 1" only include one content: the existence theorem of solutions to differential equation(s). 
This is a theorem which is needed to be proved, not an axiom. For  differential equations, 
the existence of solutions requires several sufficient conditions;
such as initial and boundary conditions, compatibility conditions, 
topological conditions and so on; all of them are sufficient conditions for the existence theorem of solutions. 

For Level 1, there is a \textbf{theorem I} :
all sufficient conditions must be {\bfseries independent} of the differential equations themselves;
especially all sufficient conditions must be {\bfseries independent} of "Level 2".

Please note: that the sufficient conditions for the existence of solutions are independent of the differential equations themselves is a normal logic, not a new discovery. The \textbf{theorem I} will be applied to the following sections.

\vspace{1em}

"Level 2" refers to all other theorems (excluding the existence theorem), which include propositions, inferences, assertions derived from differential equations.
All the contents of Level 2 need to be proved, the second level is built on level 0 and level 1.
Without the first two levels, level 2 does not exist. 
For example, if a differential equation has no solution, then we can't get any meaningful "theorem" from non-solutions equations.

\section{Electromagnetic Part}
First, the pure mathematical explanation is written, 
second, the explanation is employed on physics. 

\subsection{Existence of Solution}\label{sec_math}
In this subsection, there is only pure mathematics, without physics.
For the sake of better understanding, we suppose $F_{kl}$ is an antisymmetric
$n$-dimenstional ($n>1$) tensor. We consider the following equation
\begin{equation}\label{max1-n}
\partial^{l}F_{kl} =J_{k}, \quad n=2,3,4,5,\cdots .
\end{equation}
$J_k$ is a known source. $J_k$'s formalization is determined by itself.
If $\partial^{k}J_{k}\neq0$ (e.g. $J_k \propto x_k$) is chosen,
Eq.\eqref{max1-n} has no solutions.
Hence, $\partial^{k}J_{k}=0$ is one of {\textbf{preconditions}}
\footnote{The word "preconditions" mean the sufficient conditions. } 
of existence about solutions of Eq.\eqref{max1-n}.
If $\partial^{k}J_{k}=0$ is 
thought as a corollary of Eq.\eqref{max1-n}
($0\equiv \partial^{k} \partial^{l}F_{kl}=\partial^{k}J_{k}
{\ \Rightarrow \ } \partial^{k}J_{k} =0$), and not be indenpendent of Eq.\eqref{max1-n};
and at the same 
time it is one of {\textbf{preconditions}} of existence about 
Eq.\eqref{max1-n}'s  solutions. It must involve circular reasoning.

Hence, $\partial^{k}J_{k}=0$ is independent of Eq.\eqref{max1-n} for any $n(>1)$.

\subsection{Application on physics}
Now, we let $n=4$  and metric signature is $+2$, and then
Eq.\eqref{max1-n} becomes Eq.\eqref{max1}.
\begin{align}
\partial^{\beta}F_{\alpha\beta}&=J_{\alpha} \label{max1}\\
\partial_{\alpha}F_{\beta\gamma} + \partial_{\beta}F_{\gamma\alpha}
+ \partial_{\gamma}F_{\alpha\beta}&=0 \label{max2}
\end{align}
The above two equations are Maxwell's equations,
and $F_{\alpha\beta}$ is antisymmetric.
In most electrodynamics textbooks, the following derivation can be referred:
from Eq.\eqref{max1}, we can get $0\equiv\partial^{\alpha}\partial^{\beta}
F_{\alpha\beta}=\partial^{\alpha}J_{\alpha} {\ {\Rightarrow}\ } \partial^{\alpha}J_{\alpha}=0$. 
The charge conservation law ($\partial^{\alpha}J_{\alpha}=0$) is derived from Maxwell's equations.

However, in \S\ref{sec_math}, we have found that: if 
$\partial^{k}J_{k}=0$ is considered as not be independent of Eq.\eqref{max1-n},
it involve circular reasoning. Therefore the derivation is wrong.
All we can say is that: the charge conservation law is independent of Maxwell's equations,
and both are compatible each other.

\section{Gravitational part}

For an $n$-dimenstional($n>2$) manifold, $R_{kl}$ is Ricci tensor, 
and $G_{kl}\equiv R_{kl} - \frac{1}{2}g_{kl}R^{j}_{{\ }j}$.
Because of Bianchi identities,
$\nabla^{k}G_{kl}\equiv 0$ are identities for any $n$.
We consider the following equation
\begin{equation}\label{eins-n}
G_{kl} = 8\pi T_{kl}, \quad n=3,4,5,\cdots .
\end{equation}
$T_{kl}$ is a known source.
If $\nabla^{k}T_{kl}\neq 0$, solutions of Eqs.\eqref{eins-n} do not exist.
Therefore, $\nabla^{k}T_{kl}= 0$  is one of {\textbf{preconditions}} 
of existence about Eqs.\eqref{eins-n}'s solutions.
If we think that $\nabla^{k}T_{kl}= 0$  not be independent of
Eqs.\eqref{eins-n}
($0\equiv \nabla^{k}G_{kl}=8\pi \nabla^{k}T_{kl}
{\ \Rightarrow \ } \nabla^{k}T_{kl} =0$), 
the circular reasoning must be involved.
Hence, $\nabla^{k}T_{kl}=0$ is independent of Eq.\eqref{eins-n} for any $n$.

In \S 5.4 of \cite{poisson}, the author thinks that $\nabla^{k}T_{kl}= 0$  are not independent of 
$G_{kl} = 8\pi T_{kl}$. However, the author has changed his viewpoint about it in \cite{will} by private communications.


The same thing also happens in Yang-Mills fields and corresponding sources.

That existing {\textbf{differential identities}} are the same 
thing between Maxwell's, Einstein's and Yang-Mills's equations.

\section{Maxwell's equations in 3+1 form}
Now, we talk the relation between Maxwell's curl equations and divergence ones.
Maxwell's equations without sources are:
\begin{align}
\nabla  \cdot  {\bf{B}} &= 0 , &  \nabla  \cdot {\bf{E}} &= 0, \label{max_div} \\
\nabla  \times {\bf{B}} &=  \dfrac{{\partial {\bf{E}}}}{{\partial t}}, &
\nabla  \times {\bf{E}} &=  - \dfrac{{\partial {\bf{B}}}}{{\partial t}} \label{max_curl}.
\end{align}

Taking the divergence of Eqs.\eqref{max_curl} gives
\begin{equation} \label{initbe}
\frac{\partial }{{\partial t}}\left( {\nabla  \cdot {\bf{E}}  } \right)=0,\quad
\frac{\partial }{{\partial t}}\left( {\nabla  \cdot {\bf{B}}  } \right)=0.
\end{equation}

Following the above analysis, that compatibility conditions Eqs.\eqref{initbe} 
hold is one of {\textbf{preconditions}} of existence about solutions 
of Eqs.\eqref{max_curl}. And that Eqs.\eqref{max_div} hold can ensure 
Eqs.\eqref{initbe} hold. Therefore, the viewpoint in Ref.{\cite{courant,stratton1941}} 
that Eqs.\eqref{max_div} are thought as initial conditions of 
Eqs.\eqref{max_curl} is not correct. If the  view of Stratton{\cite{courant,stratton1941}} 
is right, a similar circular reasoning exists in the derivation.


\section{Overdetermination}
Now, we  talk the last thing. Maxwell's equations 
(Eqs.\eqref{max1},\eqref{max2})/\eqref{max_div},\eqref{max_curl}), 
Einstein's equations (with four harmonic coordinates) and 
Yang-Mills's equations (with gauge conditions) are overdetermined systems. 
A generalized definition can be employed to describe the overdetermination. 
There are first-order linear partial differential equations as following
\begin{equation} \label{lde1}
\left\{ {\begin{array}{*{20}{c}}
    \sum\limits_{ij}^{} {a_{ij}^{( 1 )}\dfrac{{\partial {y_j}}}{{\partial {x_i}}}}  + {f_1} = 0  \\
    \vdots \\
    \sum\limits_{ij}^{} {a_{ij}^{( n )}\dfrac{{\partial {y_j}}}{{\partial {x_i}}}}  + {f_n} = 0
    \end{array}} \right.
\end{equation}
where $x_i$ are independent variables; $y_j$ are dependent unknowns; 
$a_{ij}^{( k )}$ are linear coefficients; 
and  $f_k$ are non-homogeneous items. 
Let  ${Z_k} \equiv \sum_{ij}^{} 
{a_{ij}^{( k )}\frac{{\partial {y_j}}}{{\partial {x_i}}}}  + {f_k}$.

Two linear dependence definitions are as following.

{\textbf{Definition I}}:  In algebra, given a number field $P$, 
when there are coefficients ($c_k\in P$), not all zero, 
such that $\sum_{k} {{c_k}{Z_k} = 0} $; the Eqs.\eqref{lde1} are linear dependent.

This definition can be referred in any algebraic textbook.
Maxwell's equations are over-determined in the definition I.

{\textbf{Definition II (differential linear dependence)}}:   
Given a number field $P$, when there are coefficients (${c_k},d_{kl}\in P$), 
not all zero, such that {{$\sum_k {{c_k}{Z_k}}  + \sum_{kl} {{d_{kl}}
            \frac{{\partial }}{{\partial {x_l}}}{Z_k} = 0}$}},
the Eqs.\eqref{lde1} are thought as \emph{differential} linear dependent. 
If ${d_{kl}} \equiv 0$, this definition degenerates into the definition I.

Maxwell's equations (Eqs.\eqref{max1},\eqref{max2}/\eqref{max_div},\eqref{max_curl}), 
Einstein's equations (with four harmonic coordinates) and 
Yang-Mills's equations (with gauge conditions) are well-determined in  {\textbf{definition II}}.

There are some unproved propositions  about the definition II.

\textcircled{1} If Eqs.{\eqref{lde1}}, whose solutions exist and are unique, 
are over-determined in the definition I, then they must be well-determined in the  definition II.

\textcircled{2} If Eqs.{\eqref{lde1}}, whose solutions exist, are under-determined 
in  the definition II and  are well-determined in the definition I, 
then the solutions must be non-unique.

\textcircled{3} If Eqs.{\eqref{lde1}} are over-determined in  the definition II, 
then the solutions do not exist.

The unproved propositions  seem obvious, but the proof is not easy. 
If all the propositions are correct, the  definition I should be changed to the definition II.




%
%



\end{document}